\documentclass[runningheads]{llncs}
\usepackage{graphicx}
\usepackage[inline]{enumitem}
\usepackage{cite}
\usepackage[strings]{underscore}
\newlist{questions}{enumerate}{2}
\setlist[questions,1]{label=RQ\arabic*.,ref=RQ\arabic*}
\setlist[questions,2]{label=(\alph*),ref=\thequestionsi(\alph*)}

\begin{document}
\title{Knowledge-based Multimodal Music Similarity}
\author{Andrea Poltronieri\inst{1}\orcidID{0000-0003-3848-7574}} 
\authorrunning{A. Poltronieri}
\institute{Department of Computer Science and Engineering, University of Bologna, Italy
\email{andrea.poltronieri2@unibo.it}}
\maketitle              % typeset the header of the contribution
\begin{abstract}
Music similarity is an essential aspect of music retrieval, recommendation systems, and music analysis. Moreover, similarity is of vital interest for music experts, as it allows studying analogies and influences among composers and historical periods.

Current approaches to musical similarity rely mainly on symbolic content, which can be expensive to produce and is not always readily available. 
%In contrast, approaches using audio signals generally do not provide any information on the reason for the similarity.
Conversely, approaches using audio signals typically fail to provide any insight about the reasons behind the observed similarity.

This research addresses the limitations of current approaches by focusing on the study of musical similarity using both symbolic and audio content. The aim of this research is to develop a fully explainable and interpretable system that can provide end-users with more control and understanding of music similarity and classification systems.

%To achieve this goal, the proposed system uses a multimodal approach. First, the symbolic content is studied to assess similarity in a transparent and explainable way. Then, the symbolic content is aligned with the audio signal using multimodal datasets. Finally, a deep learning system is trained to analyse the audio signal informed by the symbolically-analysed similarities. 
\keywords{Music Similarity  \and Computational Musicology \and Knowledge Graphs.}
\end{abstract}

\section{Introduction}

%Music is a fundamental aspect of human culture, and its study has been explored across various disciplines, such as psychology, sociology, and musicology. Among these areas, the research of music similarity and classification has been gaining attention in recent years, in both research and commercial contexts \cite{Knees2016}. 
% Understanding and classifying music according to its similarity can have a plethora of applications, from music recommendation systems to music cognition. The primary objective of this thesis proposal is to investigate the current state of research in music similarity and classification and to propose a novel approach for understanding and classifying music.

Music similarity is a central area of research in the field of Music Information Retrieval (MIR) \cite{downie2004mir} as it enables various applications, such as music recommendation, playlist generation, music search, and classification. The ability to measure the similarity between music tracks is essential for providing personalised and relevant recommendations to users based on their listening history and preferences \cite{mcfee2012similarityrecommendation}. 
Music similarity also facilitates the discovery of new music that matches the user's taste \cite{mehrotra2021discovery}. Additionally, music similarity can be used for content-based music classification, such as genre classification \cite{correa2016survey}. 
It is also useful in musicological research, as it allows for the exploration of musical patterns and structures across different styles and genres \cite{velardo2016}. 

\subsection{Problem Statement}

The study of musical similarity is approached from various perspectives, which can be summarised in \emph{content-based systems} and \emph{context-based systems} \cite{knees2013survey}. 
The former approach extracts information directly from the musical content (whether symbolic or audio), while the latter obtains information from non-musical data, such as metadata or information related to the song's popularity or listener characteristics. 
%The primary focus of this research is on content-based similarity, as it proposes a measure of similarity that can be based on factual data and quantitatively measured, such as harmonic and melodic information.
Content-based approaches allow a quantitative measurement of similarity based on factual music data, and make it possible to investigate similarities independently of the availability and accuracy of metadata \cite{Karydis2013content}. 

However, studying content-based similarity poses several challenges, given the multidisciplinary nature of the research, which encompasses music theory, ethnomusicology, cognitive science, and computer science \cite{velardo2016}.

In content-based music similarity, a further distinction must be made concerning the representation of music. Two types of representations have been identified: \emph{signal representations} that are recordings of sound sources, and \emph{symbolic representations} that represent discrete musical events \cite{vinet2003levels}. 
Symbolic representations are context-aware and offer a structured representation from which is easy to extract information from. On the other hand, signal representations are content-unaware and not structured, which makes  extracting information from them a challenging task \cite{wiggins1993representation}.
Signal representations are by far more studied than symbolic representations, since they are more interesting from a commercial point of view (e.g. for streaming services) and the data availability is higher.

Depending on the type of musical representation, several features can be used for similarity analysis: \emph{descriptive metadata}, \emph{low-level features}, and \emph{high-level features} \cite{zheng2017genre}. Descriptive metadata is text-based information about the song, while low-level features are extracted from the audio signal (e.g., beat, tempo) and are efficient but difficult to interpret. High-level features, on the other hand, are content descriptors that reflect the knowledge of experienced or professional listeners, making them the most intuitive approach for music classification tasks.

Most of the available music similarity systems, especially those based on audio signals \cite{velardo2016}, rely on low-level features. Annotating high-level content descriptors is also expensive and requires the expertise of musicians and musicologists \cite{hao2020feature}. As a result, most available systems cannot explicitly recognise similarity motives, and their lack of interpretability and transparency can lead to biased recommendations.

%Nonetheless, transparency in these systems is essential for improving end-user control, increasing acceptance of complex algorithmic systems, and promoting user learning from the data \cite{kulesza2015personalize}. 
This results in a measure of similarity that is neither interpretable nor transparent, which may result in biased results \cite{kowald2020poularitybias}.
%The lack of interpretability and transparency in the computation of musical similarity has significant implications for recommendation systems, which may not provide transparent and unbiased recommendations to end-users, even when based on music content analysis (i.e., low-level audio features) \cite{ribeiro2016recommendation}. Such systems may carry bias in terms of the popularity of the song \cite{kowald2020poularitybias}, the personality characteristics of the end user \cite{melchiorre2020personalitybias}, and its genre \cite{shakespeare2020genrebias}.

%These problems lead to the definition of the research questions that drive this study: 

%\begin{itemize}[label={}]
%    \item \textbf{RQ1.}
%    \item \textbf{RQ2.}
%\end{itemize}

\subsection{Expected Contribution}

This research proposes a fully explainable and interpretable system that provides information on musical similarity based on both symbolic and audio content, with a focus on factual musical data such as melodic and harmonic patterns. 

\begin{description}
    \item[RQ1] \textit{What is an effective method to create high-quality datasets that incorporate multimodal data that links symbolic annotations (both melodic and harmonic) and audio?}
\end{description}

To achieve this, the symbolic content needs to be studied first to assess similarity in a transparent and explainable way. 

\begin{description}
    \item[RQ2] \textit{How can similarity measures be derived from this knowledge graph in order for it to be objectively measured and quantified?}
\end{description}

Next, an alignment of the symbolic content with the audio signal using multimodal datasets must be performed. Finally, a deep learning system is trained to analyse the audio signal informed by the symbolic content. 
By doing so, it is possible to provide end-users with more control and understanding of the music similarity and classification systems they use, regardless of the representation under analysis.

\begin{description}
    \item[RQ3] \textit{How can score-informed audio analysis be used to identify similarities and patterns in audio data, and what are the benefits of this approach for the study of music similarity?}
\end{description}

The current study focuses on the application of Semantic Web technologies, particularly in the representation and alignment of multimodal data. One of the key challenges is how to effectively encode knowledge graphs (KGs) to enable their use as input and mapping onto various mathematical models, such as timeseries and embedding.

\section{Related Works}

%Music similarity and classification are complex tasks that have been extensively studied in recent years. However, despite the different approaches and techniques developed, there are still several limitations that need to be addressed. These limitations are discussed according to the source analysed: symbolic music and audio signals.
%Furthermore, in this section, the strengths and limitations of multimodal music similarity are presented and discussed.

% In conclusion, despite the progress made in recent years, music similarity and classification still presents several limitations. These limitations include a lack of meaningful similarity measures in the symbolic music domain, a lack of attention given to harmonic similarity, and a lack of comprehensive analysis of the audio features in the audio signal domain.

\subsection{Symbolic Music Similarity}

The study of similarity on symbolic content has been studied in depth in recent years. Various approaches have been proposed, ranging from harmonic similarity to melodic and rhythmic similarity. 
%The study of symbolic similarity has been applied to a large number of tasks, such as genre classification \cite{goienetxea2018similaritygenre}, plagiarism detection \cite{wolf2012plagiarism}, and style recognition \cite{ens2020style}.

Melodic similarity is the most extensively researched category. Algorithms that handle melodic similarity in symbolic form are typically rule-based and aim to define various types of context-dependent similarity functions, which rely on music theory \cite{orio2009graph}. 
However, these algorithms lack a shared definition of similarity and primarily focus on studying similarity in monophonic sequences \cite{velardo2016}.

%Recently, machine learning and deep learning-based algorithms have also been proposed for symbolic similarity \cite{karsdorp2019}. These algorithms are typically concerned with studying a generic type of similarity and do not provide any indication of the reasons for similarity between two pieces.

% Similarly, systems analyzing similarity and recommendation systems based on the audio signal are almost solely based on deep learning approaches (e.g. \cite{KULKARNI2020reccomendation, knees2013survey, gurjar2018survey}).

On the other hand, algorithms for harmonic similarity has not received much attention in recent years. To the best of my knowledge, current state-of-the-art methods for this task are the \emph{Tonal Pitch Step Distance} (TPSD) \cite{deHaas2013tpsd} and the \emph{Chord Sequence Alignment System} (CSAS) \cite{hanna2009alignment}. These studies consider tracks similar only if their harmonic profiles are globally aligned, providing no information on local similarity.

Studies using a combination of harmonic and melodic content to calculate similarity are limited to a few contributions \cite{giraud2015melodicharmonic}.

\subsection{Audio Music Similarity}

Music similarity in the audio signal domain has been studied for a wide range of applications, ranging from cover song identification \cite{SheikhFathollahi2021audiorecommendation} to recommendation systems \cite{du2022bytecover2}. 
These algorithms are based on the extraction of low-level features directly from the signal, such as spectrograms, MFCCs and Chorma Features \cite{dorfler2017spectre}.

One of the main limitations of these approaches is their reliance on deep learning approaches. These methods are based on end-to-end algorithms that do not provide valuable information regarding fundamental aspects of similarity, such as the explanation for why two or more tracks are similar, and the highlight of parts in common between different tracks. 

% Additionally, deep learning approaches require large amounts of data for training and are prone to overfitting, which can be problematic for music datasets that are limited in size.

%Another limitation in the audio signal domain is the lack of a comprehensive analysis of the audio features. Most current systems extract a limited set of features, such as tempo and melody, which may not provide a comprehensive understanding of the musical content \cite{vinet2003levels}.

\subsection{Multimodal Music Similarity}

Multimodality refers to the integration of multiple representation modes, such as visual, auditory, and textual.

%The use of multimodal architectures has been tested in several areas, with great results \cite{joshi2021mutimodal}. 
%For instance, in image captioning \cite{hossain2018captioning} and visual question answering  \cite{we2016visualqa} tasks, combining textual and visual data can provide more context and generate more informative descriptions. Similarly, incorporating physiological signals along with video and text data can lead to better sentiment analysis \cite{falk2017sentiment}. 

In the realm of music, multimodality has become an increasingly popular field of research in recent years and has proven to provide better results in different tasks, if compared to approaches that consider a single modality \cite{baltrusaitis2017multimodalml, simonetta2019multimodal}.

One of the primary areas of research in multimodal MIR is the integration of audio and textual data. 
%This involves developing algorithms that can analyse both the audio and textual components of music recordings to extract useful information, such as genre, tempo, and key \cite{manco2022lyricmulti, zhand2022lyricmulti, eileen2022lyricsmulti}.
Moreover, multimodality has been explored also for other tasks, such as audio-to-score alignment \cite{muller2015fundamentals} and classification \cite{laurier2008multimodalclassification}. 

%Multimodal algorithms have been employed for musical similarity and classification \cite{laurier2008multimodalclassification}, as well as for recommendation systems \cite{huang2020multimodalrecommendation}. 
However, less emphasis has been placed on algorithms that combine audio and symbolic annotations, particularly in the field of classification and similarity. Some methods, like \cite{Balke2016multimodalthemes} and \cite{Suyoto2008multimodalqueries}, aim to identify audio tracks through symbolic queries, but they rely on converting either audio into symbolic or symbolic into audio, respectively.
In contrast, \cite{Li2015multimodalviolin} proposes a score-informed analysis of audio. Although this approach represents a promising development, it has to be considered a preliminary study, with a small sample size of only 20 violin-only tracks.

\section{Research Metodology}

The primary objective of this research is to develop algorithms that can accurately measure musical similarity based on both audio and symbolic content. The proposed approach will consider factual musical data and provide an interpretable model for computing music similarity between music pieces.

\subsubsection{Dataset creation.}

To achieve this goal, the first step is to create a multimodal dataset, which includes various types of data for each song in the dataset (c.f. \emph{RQ1}). Specifically, the dataset must consist of four key elements for each track: 
\begin{enumerate*}[label=(\roman*)]
    \item an audio track,
    \item melodic annotations,
    \item harmonic annotations, and
    \item track metadata.
\end{enumerate*}

%This will lay the foundation for developing and training algorithms that can accurately capture musical similarity.
The dataset will be encoded as a RDF/OWL Knowledge Graph (KG) \cite{bizer2009lod}, which will define semantic relationships between the various multimodal elements. 
The KG will also contain alignment data between different types of annotations, such as audio, melodic and harmonic data.

\subsubsection{Similarity computation.}

Similarity measures based on symbolic data will then be defined (c.f. \emph{RQ2}), focusing on both melodic and harmonic elements. 
To achieve this, it is first necessary to define the concept of music similarity both musicologically and perceptually. 
First, repositories and datasets of known patterns will serve as a basis for the definition of similarity functions. 
Then, various types of matches, such as exact and fuzzy matches, will be considered between symbolic annotations at different levels, such as phrases, form, cadences, and melodies.
This approach enables the investigation of musical similarity from a purely musical perspective, which would allow to the resulting similarity functions to be both explainable and transparent. 
%Harmonic and melodic similarity will be calculated using algorithms specifically designed for this purpose, ensuring they are musically relevant and perceptually meaningful. 
%This will be achieved by using datasets of musicologically validated patterns \cite{medina2003pattern, adegbija23jazznet} and user-based studies for evaluating the algorithms.

The research will also enable the definition of local similarities, allowing for the analysis of influences between different songs, as well as the detection of plagiarism in specific song sections.
Moreover, the similarity analysis will be conducted by jointly analysing the harmonic and melodic data to provide more realistic and musicologically grounded similarity information.

\subsubsection{Multimodal analysis.}

In the final step, the similarities extracted from the symbolic data will be used to study similarity on the audio signal. This will involve training deep learning architectures on the aligned audio and symbolic data through the application of data fusion techniques (c.f. \emph{RQ3}).
Great care will be given in selecting an architecture that is both explainable and allows for analogies to be drawn between the various components of the multimodal analysis, such as deep learning architectures and neuro-symbolic reasoning. 

An architecture that will be explored is transformers \cite{chefer2020transformer}, which in this context can be employed for the unsupervised matching between symbolic annotations and audio features. 
Hence, the produced unsupervised model will be fine-tuned using the similarity measures extracted from the symbolic annotations. 

%Finally, I plan to employ the similarity measurements obtained from both the symbolic and audio data to implement classification algorithms. These algorithms will help to categorise the data and allow to draw meaningful conclusions from the performed multimodal analysis.

\subsection{Evaluation}

The validation of the results obtained will focus on two main elements: (i) similarity measures based on symbolic content; and (ii) similarity based on audio signals.

Firstly, the similarity measures calculated on symbolic content will be evaluated to determine if the output of the defined similarity functions produces a musicologically or perceptually relevant output. 
Moreover, known pattern datasets \cite{medina2003pattern, adegbija23jazznet} will be used to evaluate the output of the similarity measures.
Secondly, crowdsourced surveys will be conducted to gather more data on the perceptual relevance of the extracted similarities.

Regarding the similarities calculated on audio signals, global results will be evaluated on typical music information retrieval tasks, such as cover song detection. For local similarities, the audio extracted similarities will be evaluated using the symbolic-aligned data.

Similarly, we will assess the transparency and explainability of the model. While the explainability of the symbolic similarity models is inherent in their design, the explainability of the model on audio signal will be evaluated by comparing the results to the aligned symbolic annotations.

\section{Current Results}

As initial contributions to the development of this research project, work was conducted on several fronts, including the creation of a dataset, the study of harmonic similarity, the embedding of harmonic annotations, and the construction of ontologies for modeling musical content.

\subsection{Dataset creation}
\label{subsec:dataset}

As the first contribution of my research, I focused on the creation of a dataset of harmonic annotations (c.f. RQ1): ChoCo, the largest available \emph{Chord Corpus} \cite{deberardinis2022choco}.
Choco is a large-scale dataset that semantically integrates harmonic data from 18 different sources in various representations and formats (Harte, Leadsheet, Roman numerals, ABC). 
The corpus leverage JAMS (JSON Annotated Music Specification) \cite{humphrey2014jams}, a popular data structure for annotations in Music Information Retrieval, to effectively represent a variety of chord-related information (chord, key, mode, etc.) in a uniform way. 
ChoCo also consists of a converter module that takes care of standardising chord annotations into a single format, the Harte Notation \cite{harte2005symbolic}. 
On top of it, a novel ontology modelling music annotations and involved entities (artists, scores, etc.) has been proposed, and a 30M triple knowledge graph\footnote{ChoCo SPARQL Endpoint: \url{https://polifonia.disi.unibo.it/choco/sparql}} has been built.

%\begin{figure}[h]
%    \centering
%    \includegraphics[width=.8\linewidth]{figures/lomir_workflow.png}
%    \caption{Overview of our data transformation workflow of ChoCo.}
%    \label{fig:lomir-worKflow}
%\end{figure}

%Figure \ref{fig:lomir-worKflow} shows the workflow used for building the corpus and converting the data.
The proposed workflow is highly scalable and enables the seamless integration of additional data types, including melodic and structural annotations. Moreover, the Knowledge Graph utilised in ChoCo facilitates the alignment of its annotations with various metadata available on the web, such as MusicBrainz\footnote{MusicBrainz: \url{https://musicbrainz.org/}} and Discogs\footnote{Discogs: \url{https://www.discogs.com/}}.

As a result, these resources provide an accurate and distinct reference point for each track, which will allow the identification of the audio recording which refers to the annotations contained in the dataset.

\subsection{Studies on Harmonic Similarity}

In accordance with the second research question (RQ2), a preliminary investigation into the similarity measures has been conducted.

Based on the limitations found in the state-of-the-art study of harmonic similarity, I worked on LHARP, a \emph{Local Harmonic Agreement of Recurrent Patterns}.
LHARP is a measure of harmonic similarity formulated for emphasising shared repeated patterns among two arbitrary symbolic sequences, thereby providing a general framework for the analysis of symbolic streams based on their local structures.

To evaluate the efficacy of LHARP as a method for harmonic similarity, two separate experiments were carried out -- each pertaining to a case study that the function can potentially accommodate.
First, a graph analysis was performed to encode harmonic dependencies (edges) between music pieces (nodes) based on their similarity values.
%\footnote{LHARP graph: \url{https://polifonia-project.github.io/musilar-preview/}}.
Second, to conform with the literature, a cover song detection experiment was conducted.

\begin{figure}[t]
    \centering
    \includegraphics[width=.8\linewidth]{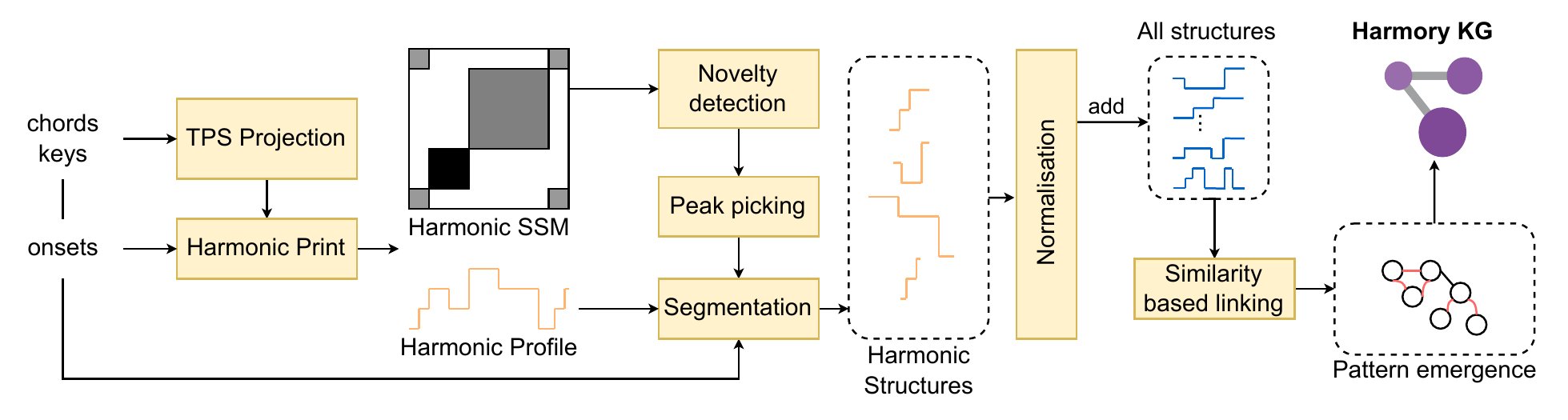}
    \caption{Workflow used for the production of the Harmonic Memory (Harmory).}
    \label{fig:harmory}
\end{figure}

As an evolution of LHARP, I worked on the \emph{Harmonic Memory} (Harmory) \cite{deberardinis2023harmory}. 
Harmory is a Knowledge Graph (KG) of harmonic patterns extracted from a large and heterogeneous musical corpus. By leveraging a cognitive model of tonal harmony, chord progressions are segmented into meaningful structures, and patterns emerge from their comparison via harmonic similarity. 
Akin to a music memory, the KG holds temporal connections between consecutive patterns, as well as salient similarity relationships (c.f. Figure \ref{fig:harmory}). 
%Harmory enables novel pathways for combinational creativity: the memory provides a fully accountable and explainable framework to inspire and support creative professionals – allowing for the discovery of progressions consistent with given criteria, the recomposition of harmonic sections, but also the co-creation of new progressions.

%Figure \ref{fig:harmory} provides a visualization of the workflow utilized for the development of Harmory.

During the creation of Harmory, I focused on the developement of both harmonic segmentation and harmonic similarity state-of-the-art algorithm.

Digital Signal Processing (DSP) algorithms were used to perform harmonic segmentation on symbolic content. Tonal Pitch Space (TPS) \cite{lerdahl1988tps} was used to encode the harmonic sequences and generate a Self-Similarity Matrix (SSM) \cite{de2020unveiling}, from which a novelty curve was extracted to identify the harmonic segment boundaries \cite{muller2015fundamentals}. 

Additionally, a new algorithm for computing harmonic similarity using Dynamic Time Warping (DTW) \cite{sakoe1978dtw} on TPS-encoded sequences was proposed, which is more efficient than the previous state-of-the-art approach \cite{deHaas2013tpsd}.

%Moreover, it establishes a new state-of-the-art benchmark for the harmonic segmentation task in symbolic music.

\subsection{Music Chord Embeddings}

Another aspect of my work involved the definition of embeddings to enable the expressive encoding of harmonic annotations. To achieve this goal, I developed \emph{pitchclass2vec}, a novel type of embedding that effectively preserves the harmonic characteristics of a chord.

The efficacy of this embedding was evaluated in a Music Structure Analysis task, where it outperformed other approaches, including those based on chord encoding \cite{madjiheurem2016chord2vec} or textual encoding \cite{bojanowski2018fasttext}.

\subsection{Semantic Integration of Musical Data}

For the development of the aforementioned works, ontologies were created to model various types of data related to the music domain. 
These works respond to RQ1, and aim to provide new methods for the representation of musical knowledge.
These ontologies include the \emph{JAMS Ontology}
%\footnote{JAMS ontology: \url{https://github.com/polifonia-project/jams-ontology}}
, which models musical notations (such as chords, patterns, and musical structures), the \emph{Roman Chord Ontology}
%\footnote{Roman Chord Ontology: \url{https://github.com/polifonia-project/roman-chord-ontology}}
, which models chords expressed in Roman numeral notation, and the \emph{Music Note Ontology}
%\footnote{Music Note Ontology: \url{https://github.com/andreamust/music_note_pattern}}
, which models musical notes and their realisation (i.e., the note played in a performance).
These ontologies are part of an ontological framework named \emph{Polifonia Ontology Network} (PON).
%\footnote{Polifonia Ontology Network: \url{https://github.com/polifonia-project/ontology-network}} \cite{carriero2021semanticintegration}. 

\section{Conclusion and Next Steps}

This paper presents a research project that employs a symbolic-informed architecture to study music similarities on audio signals. 
This allows an explainable and interpretable musically-grounded analysis of similarities in music which can be performed both on symbolic annotations and audio signal.
%This achievement establishes a framework for exploring the applications of similarity within an empirical and explicable context.

The use of Knowledge Graphs (KG) and Semantic Web tools is crucial to this research as they provide a foundation for data alignment and interoperability across various data types.

Moving forward, the research will focus on expanding the dataset (as described in Section \ref{subsec:dataset}) by incorporating new data types, such as melodic data and audio signals, into the knowledge graph. This will facilitate exploration of novel similarity functions that enable the study of symbolic data, integrating diverse musical elements such as melody, harmony, and structure.

Subsequently, the research will aim to align the produced data with audio signals, with the objective of training a model informed by symbolic data that is capable of analysing similarity on audio signals.

Finally, a crucial objective of this study is to extend the ontological models developed to enable multimodal analysis of other data types and in other domains.

\section*{Acknowledgements}
This project has received funding from the European Union's Horizon 2020 research and innovation programme under grant agreement No 101004746.
Moreover, I would like to express my gratitude to Prof. Valentina Presutti, my supervisor, for her guidance and contributions to this research project. I also extend my thanks to Prof Victor de Boer, my mentor for the Doctoral Symposium, for sharing valuable insights and advice.
\bibliographystyle{splncs04}
\bibliography{bibliography}

\begin{thebibliography}{10}
\providecommand{\url}[1]{\texttt{#1}}
\providecommand{\urlprefix}{URL }
\providecommand{\doi}[1]{https://doi.org/#1}

\bibitem{adegbija23jazznet}
Adegbija, T.: jazznet: A dataset of fundamental piano patterns for music audio
  machine learning research. In: IEEE International Conference on Acoustics,
  Speech and Signal Processing (ICASSP). IEEE (2023)

\bibitem{Balke2016multimodalthemes}
Balke, S., Arifi-M{\"u}ller, V., Lamprecht, L., M{\"u}ller, M.: Retrieving
  audio recordings using musical themes. In: 2016 {IEEE} International
  Conference on Acoustics, Speech and Signal Processing ({ICASSP}). pp.
  281--285 (Mar 2016)

\bibitem{baltrusaitis2017multimodalml}
Baltrušaitis, T., Ahuja, C., Morency, L.P.: Multimodal machine learning: A
  survey and taxonomy (2017)

\bibitem{deberardinis2022choco}
de~Berardinis, J., Meroño-Peñuela, A., Poltronieri, A., Presutti, V.: Choco:
  a chord corpus and a data transformation workflow for musical harmony
  knowledge graphs. In: Manuscript under review (2022)

\bibitem{deberardinis2023harmory}
de~Berardinis, J., Meroño-Peñuela, A., Poltronieri, A., Presutti, V.: The
  harmonic memory: a knowledge graph of harmonic patterns as a trustworthy
  framework for computational creativity. In: The Web Conference, to be
  published (2023)

\bibitem{de2020unveiling}
de~Berardinis, J., Vamvakaris, M., Cangelosi, A., Coutinho, E.: Unveiling the
  hierarchical structure of music by multi-resolution community detection.
  Transactions of the International Society for Music Information Retrieval
  \textbf{3}(1),  82--97 (2020)

\bibitem{bizer2009lod}
Bizer, C., Heath, T., Berners{-}Lee, T.: Linked data - the story so far. Int.
  J. Semantic Web Inf. Syst.  \textbf{5}(3),  1--22 (2009)

\bibitem{bojanowski2018fasttext}
Bojanowski, P., Grave, E., Joulin, A., Mikolov, T.: Enriching word vectors with
  subword information. Trans. Assoc. Comput. Linguistics  \textbf{5},  135--146
  (2017)

\bibitem{chefer2020transformer}
Chefer, H., Gur, S., Wolf, L.: Transformer interpretability beyond attention
  visualization (2020)

\bibitem{correa2016survey}
Corrêa, D.C., Rodrigues, F.A.: A survey on symbolic data-based music genre
  classification. Expert Systems with Applications  \textbf{60},  190--210
  (2016)

\bibitem{downie2004mir}
Downie, J.S.: The scientific evaluation of music information retrieval systems:
  Foundations and future. Comput. Music. J.  \textbf{28}(2),  12--23 (2004)

\bibitem{du2022bytecover2}
Du, X., Chen, K., Wang, Z., Zhu, B., Ma, Z.: Bytecover2: Towards dimensionality
  reduction of latent embedding for efficient cover song identification. In:
  ICASSP 2022 - 2022 IEEE International Conference on Acoustics, Speech and
  Signal Processing (ICASSP). pp. 616--620 (2022)

\bibitem{dorfler2017spectre}
Dörfler, M., Bammer, R., Grill, T.: Inside the spectrogram: Convolutional
  neural networks in audio processing. In: 2017 International Conference on
  Sampling Theory and Applications (SampTA). pp. 152--155 (2017)

\bibitem{giraud2015melodicharmonic}
Giraud, M., Groult, R., Leguy, E., Lev{\'e}, F.: {Computational Fugue
  Analysis}. {Computer Music Journal}  \textbf{39}(2),  77--96 (2015)

\bibitem{deHaas2013tpsd}
de~Haas, W.B., Wiering, F., Veltkamp, R.C.: A geometrical distance measure for
  determining the similarity of musical harmony. International Journal of
  Multimedia Information Retrieval  \textbf{2}(3),  189--202 (Sep 2013)

\bibitem{hanna2009alignment}
Hanna, P., Robine, M., Rocher, T.: An alignment based system for chord sequence
  retrieval. In: Proceedings of the 9th ACM/IEEE-CS joint conference on Digital
  libraries. pp. 101--104 (2009)

\bibitem{harte2005symbolic}
Harte, C., Sandler, M.B., Abdallah, S.A., G{\'o}mez, E.: Symbolic
  representation of musical chords: A proposed syntax for text annotations. In:
  ISMIR. vol.~5, pp. 66--71 (2005)

\bibitem{humphrey2014jams}
Humphrey, E.J., Salamon, J., Nieto, O., Forsyth, J., Bittner, R.M., Bello,
  J.P.: {JAMS: A JSON Annotated Music Specification for Reproducible MIR
  Research.} In: ISMIR. pp. 591--596 (2014)

\bibitem{Karydis2013content}
Karydis, I., {Lida Kermanidis}, K., Sioutas, S., Iliadis, L.: Comparing content
  and context based similarity for musical data. Neurocomputing  \textbf{107},
  69--76 (2013), timely Neural Networks Applications in Engineering

\bibitem{knees2013survey}
Knees, P., Schedl, M.: A survey of music similarity and recommendation from
  music context data. ACM Trans. Multimedia Comput. Commun. Appl.
  \textbf{10}(1) (dec 2013)

\bibitem{kowald2020poularitybias}
Kowald, D., Schedl, M., Lex, E.: The unfairness of popularity bias in music
  recommendation: A reproducibility study. In: Jose, J.M., Yilmaz, E.,
  Magalh{\~a}es, J., Castells, P., Ferro, N., Silva, M.J., Martins, F. (eds.)
  Advances in Information Retrieval. pp. 35--42. Springer International
  Publishing, Cham (2020)

\bibitem{laurier2008multimodalclassification}
Laurier, C., Grivolla, J., Herrera, P.: Multimodal music mood classification
  using audio and lyrics. In: 2008 Seventh International Conference on Machine
  Learning and Applications. pp. 688--693 (2008)

\bibitem{lerdahl1988tps}
Lerdahl, F.: Tonal pitch space. Music Perception: An Interdisciplinary Journal
  \textbf{5}(3),  315--349 (1988)

\bibitem{Li2015multimodalviolin}
Li, P.C., Su, L., Yang, Y.H., Su, A.W.Y.: Analysis of expressive musical terms
  in violin using score-informed and expression-based audio features. In:
  International Society for Music Information Retrieval Conference (2015)

\bibitem{madjiheurem2016chord2vec}
Madjiheurem, S., Qu, L., Walder, C.: Chord2vec: Learning musical chord
  embeddings. In: Proceedings of the constructive machine learning workshop at
  30th conference on neural information processing systems (NIPS2016),
  Barcelona, Spain (2016)

\bibitem{mcfee2012similarityrecommendation}
McFee, B., Barrington, L., Lanckriet, G.: Learning content similarity for music
  recommendation. IEEE Transactions on Audio, Speech, and Language Processing
  \textbf{20}(8),  2207--2218 (2012)

\bibitem{medina2003pattern}
Medina, R., Smith, L., Wagner, D.: Content-based indexing of musical scores.
  In: 2003 Joint Conference on Digital Libraries, 2003. Proceedings. pp. 18--26
  (2003)

\bibitem{mehrotra2021discovery}
Mehrotra, R.: Algorithmic balancing of familiarity, similarity, \& discovery in
  music recommendations. In: Proceedings of the 30th ACM International
  Conference on Information \& Knowledge Management. p. 3996–4005. CIKM '21,
  Association for Computing Machinery, New York, NY, USA (2021)

\bibitem{muller2015fundamentals}
M{\"u}ller, M.: Fundamentals of music processing: Audio, analysis, algorithms,
  applications, vol.~5. Springer (2015)

\bibitem{orio2009graph}
Orio, N., Rod{\`{a}}, A.: A measure of melodic similarity based on a graph
  representation of the music structure. In: Hirata, K., Tzanetakis, G.,
  Yoshii, K. (eds.) Proceedings of the 10th International Society for Music
  Information Retrieval Conference, {ISMIR} 2009, Kobe International Conference
  Center, Kobe, Japan, October 26-30, 2009. pp. 543--548. International Society
  for Music Information Retrieval (2009)

\bibitem{sakoe1978dtw}
Sakoe, H., Chiba, S.: Dynamic programming algorithm optimization for spoken
  word recognition. IEEE Transactions on Acoustics, Speech, and Signal
  Processing  \textbf{26}(1),  43--49 (1978)

\bibitem{SheikhFathollahi2021audiorecommendation}
Sheikh~Fathollahi, M., Razzazi, F.: Music similarity measurement and
  recommendation system using convolutional neural networks. International
  Journal of Multimedia Information Retrieval  \textbf{10}(1),  43--53 (Mar
  2021)

\bibitem{simonetta2019multimodal}
Simonetta, F., Ntalampiras, S., Avanzini, F.: Multimodal music information
  processing and retrieval: Survey and future challenges. In: 2019
  International Workshop on Multilayer Music Representation and Processing
  (MMRP). pp. 10--18 (2019)

\bibitem{Suyoto2008multimodalqueries}
Suyoto, I.S.H., Uitdenbogerd, A.L., Scholer, F.: Searching musical audio using
  symbolic queries. IEEE Trans. Audio Speech Lang. Processing  \textbf{16}(2),
  372--381 (Feb 2008)

\bibitem{hao2020feature}
Tan, H.H., Herremans, D.: Music fadernets: Controllable music generation based
  on high-level features via low-level feature modelling. In: Cumming, J., Lee,
  J.H., McFee, B., Schedl, M., Devaney, J., McKay, C., Zangerle, E., de~Reuse,
  T. (eds.) Proceedings of the 21th International Society for Music Information
  Retrieval Conference, {ISMIR} 2020, Montreal, Canada, October 11-16, 2020.
  pp. 109--116 (2020)

\bibitem{velardo2016}
Velardo, V., Vallati, M., Jan, S.: {Symbolic Melodic Similarity: State of the
  Art and Future Challenges}. Computer Music Journal  \textbf{40}(2),  70--83
  (06 2016)

\bibitem{vinet2003levels}
Vinet, H.: The representation levels of music information. In: Wiil, U.K. (ed.)
  Computer Music Modeling and Retrieval. pp. 193--209. Springer Berlin
  Heidelberg, Berlin, Heidelberg (2004)

\bibitem{wiggins1993representation}
Wiggins, G., Miranda, E., Smaill, A., Harris, M.: A framework for the
  evaluation of music representation systems. Computer Music Journal
  \textbf{17}(3),  31--42 (1993)

\bibitem{zheng2017genre}
Zheng, E., Moh, M., Moh, T.S.: Music genre classification: A n-gram based
  musicological approach. In: 2017 IEEE 7th International Advance Computing
  Conference (IACC). pp. 671--677 (2017)

\end{thebibliography}
\end{document}